# Understanding Phishers' Strategies of Mimicking URLs to Leverage Phishing Attacks: A Machine Learning Approach

J. Samantha Tharani*[1] | Nalin A.G. Arachchilage[2,1]

[1]Department of Computer Science, University of Jaffna, Sri Lanaka
[2]Department of Computer Science and IT, School of Engineering and Mathematical Sciences
  La Trobe University, Australia

**Correspondence**
*Email: samantha@univ.jfn.ac.lk

**Abstract**

Phishing is a type of social engineering attack with an intention to steal user data, including login credentials and credit card numbers, leading to financial losses for both organisations and individuals. It occurs when an attacker, pretending as a trusted entity, lure a victim into click on a link or attachment in an email, or in a text message. Phishing is often launched via email messages or text messages over social networks. Previous research has revealed that phishing attacks can be identified just by looking at URLs. Identifying the techniques which are used by phishers to mimic a phishing URL is rather a challenging issue. At present, we have limited knowledge and understanding of how cyber-criminals attempt to mimic URLs with the same look and feel of the legitimate ones, to entice people into clicking links. Therefore, this paper investigates the feature selection of phishing URLs (Uniform Resource Locators), aiming to explore the strategies employed by phishers to mimic URLs that can obviously trick people into clicking links. We employed an Information Gain (IG) and Chi-Squared feature selection methods in Machine Learning (ML) on a phishing dataset. The dataset contains a total of 48 features extracted from 5000 phishing and another 5000 legitimate URL from web pages downloaded from January to May 2015 and from May to June 2017. Our results revealed that there were 10 techniques that phishers used to mimic URLs to manipulate humans into clicking links. Identifying these phishing URL manipulation techniques would certainly help to educate individuals and organisations and keep them safe from phishing attacks. In addition, the findings of this research will also help develop anti-phishing tools, framework or browser plugins for phishing prevention.

## 1  INTRODUCTION

Internet technology is so pervasive today, as it provides a baseline for people to do online banking, education, entertainment and social networking [1]. This opens up the back door for cyber-criminals to hack into sensitive information of individuals or organisation [2]. It often happens through social engineering (i.e. phishing) [3].

Phishing is a type of semantic attack [4], often used to steal user sensitive information including login credentials and credit card numbers [5][6][7]. It occurs when an attacker, masquerading as a trusted entity, entice a victim into clicking on a link or opening an attachment in an email or instant message through social messaging services such as WhatsApp, Viber or Facebook Messenger[8]. Victim is then tricked into clicking a malicious link or open an attachment, perhaps leading to install malware (i.e. malicious IT application), which can disturb the normal behaviour of a computer system[6]. For example, once the phishing link is clicked, it can automatically download a malicious IT application, so called "ransomware", which encrypts all the files, folders, images, videos, and audios on the victim's computer system[9].

Detecting phishing attacks is a challenging task [10], due to methods used by phishers to trick people into clicking links. They often employ various techniques to mimic phishing web addresses, so called Unified Resource Locations (URLs), to fool the users. For example, domain name in the URL is unknown or misspelled - domain name in the phishing URL is different to the legitimate organisation. This can often be an unknown domain or rather a misspelled version of the legitimate domain such as http://www.g0og1e.com (a.k.a. typo squatting [11]). There is another popular method is Internationalised Domain Name (IDN) spoofing or homograph attack. In this, phishers buy a URL that includes characters that appear to be English letters, but are actually from a different language set. For example Latin "c" or "a" being replaced by the Cyrillic "c" or "a". Another well known technique is Open URL Redirection[12], where a malicious script is tacked onto what appears to be a legitimate website address. But it takes the visitor to a phishing website without the users knowledge. However, they have used a set of common features like Lengthy URL, Anchor URL, Suffix and Prefix special characters, Irregular URL, etc [13] to mimic these kind of phishing URLs' techniques.

In general, there are two types of methods stated in the past research to detect and protect people from phishing attacks [14]. First one is based on blacklist, by comparing the requested URL with the URLs available in the blacklist. The downside of this approach is that the blacklist usually covers all phishing websites, nevertheless a new phishing website appears in a short while [15]. The second one is heuristic-based approach. In this, several features are extracted from the phishing URL to classify either fraudulent or legitimate. In the past literature Waleed Ali [16] proposed a phishing detection approach based on supervised learning with wrapper features selection, Mofleh-Al-diabat[13] used classification mining techniques for detecting and predicting phishing website, Neda Abdelhamid [17] proposed an approach for phishing detection based on the associative classification and Rami M. Mohammed[18] used Self-structuring Neural Network for predicting phishing website. However, previous research has failed to identify the techniques or strategies used by cyber-criminals (i.e. phishers) to mimic legitimate URLs [19]. Identifying techniques or strategies used by cyber-criminals (i.e. phishers) to mimic phishing URL is imperative to combat against phishing attacks [3] (i.e. developing anti-phishing tools, frameworks and browser plugins development).

Therefore, this research focuses on investigating the feature selection of phishing URLs (Uniform Resource Locators), aiming to explore the techniques employed by phishers to mimic phishing URLs that can obviously trick people into clicking links. Machine Learning (ML) techniques enable in classifying various features of phishing URLs [20][11]. Algorithms provided in ML help identify the pattern of phishing URL, sentence form of the phishing emails, identify suspicious attachments and links in the phishing email, and even measure the emotional factor of the phishing email, etc. In this research, we employ ML feature selection techniques (i.e. Information Gain (IG) and Chi-Squared) around 10000 URLs to identify specific features of how phishers mimic URLs to leverage their attack. The data-set used in this study collected from the [21] website that were collected from January to May 2015 and from May to June 2017.

The rest of the paper is organised as follows. Section 2 discusses related works and compare different phishing feature extraction methods presented in the literature.Section 3 describes the methodology for feature selection of phishing URLs, aiming to explore the techniques used by phishers to mimic legitimate URLs. Section 4 presents the results and discusses the uniqueness of the proposed research work. Finally, we conclude the paper in Section 5 opening up for future work.

## 2   RELATED WORK

Walled Ali[16] proposed a wrapper-based feature selection method to select the most significant features in predicting the phishing websites accurately.Authors have used a phishing website, dataset derived from UCI Machine Learning Repository[22]. Their findings revealed that supervised learning classifiers, Back-Propagation Neural Networks (BPNN), k-Nearest Neighbours (KNN), and Random-Forest (RF) in ML are achieving the best Correct Classification Rate (CCR)[11] and Radial Basis Function Network (RBFN), Naive Bayes (NB) achieved the worst CCR for detecting phishing websites. The authors revealed 30 key features of phishing websites through the dataset. The best feature subset is decided based on the highest evaluation to be used in the training of the machine learning classifiers. The downside of this feature selection approach is, consume more time and requires extra computational overhead with some classifiers. Dataset they are used for the research is pretty old [22] and most of the identified features are not only URL-related, but also domain-related and content-related. Those features haven't revealed the modern techniques that are used to mimic URLs such as Null Self Redirect Hyperlinks (NSRH) in URL, Domain name mismatch, Number of dashes in the URL, etc [23]. Furthermore, the authors failed to articulate how cyber-criminals used aforementioned features to manipulate people through phishing attacks.

Jain, Ankit Kumar and Gupta, Brij B [24] presented a survey report on phishing detection approaches based on visual similarity. They have provided a feature set of visual similarity like text content, text format, HTML tags, Cascading Style Sheet (CSS), image and so forth. It provides a better understanding of phishing website, various solutions, and future scope in phishing detection. In addition, they have pointed out the limitations in phishing detection like accuracy, the countermeasure against new phishing websites, failing to detect embedded objects. They have identified text based similarity approaches which are relatively fast, but they are unable to detect phishing attack if text is replaced with some images. Image processing based approaches has a high accuracy rate while they are complex in nature and time-consuming. Nevertheless, the authors failed to discuss the strategies or techniques used to mimic URLs.

Al-diabat, Mofleh [13] proposed a feature selection approach aiming to determine the effective set of features in-terms of improving the performance of the classification. In this research, they have used two feature selection methods such as Information Gain(IG) and Symmetrical Uncertainty (SU) to detect a small set of correlation among features. From that, they have identified eleven common features. Among those features SSL-Final-state and URL-of-Anchor identified as first two top scored features. Mined features guide the IREP and C4.5 data mining algorithms to classify phishing website with high accuracy. The labelled (-1 (Phishy) or 1 (Legitimate)) dataset contains 11000 URL samples and 30 phishing website related features collected from Phishtank [23] and Yahoo Directory [25] websites. Their result not discussed what are URL features have correlation and how they are influenced in the phishing attack or mimic URLs. Also, they are not presented how these mined features improve the classification of phishing or legitimate. If the classifier only used correlated features for the classification it might have possibility of missed some other important phishing related features.

Abdelhamid, Neda and Ayesh, Aladdin and Thabtah, Fadi [17] proposed a research work for phishing detection based on Associative Classification (AC). Their research work mainly focusing on developing rules to identify the phishing websites. Chi-square method used for feature selection to identify significant features related to the phishing website and AC data mining technique to discover the correlations among features and produces them in simple rules. Their method can able to discover new rules that are connected with more than one target class. They have achieved higher predictive accuracy by using Multi Class Classification based Association Rules(MCAR) algorithm. Its ability not only to extract one class per rule, but also all possible classes in a dis-junction form. The identified rules are inaccurate with the new features available in the latest phishing dataset[23]. Also, they are not mentioned how this approach going to incorporate if a new feature introduced by the cyber-criminals to mimic a phishing attack.

Mohammad, Rami M and Thabtah, Fadi and McCluskey, Lee [14] is mainly focused on identifying groups of features and developing a set of rules which are used to distinguish a phishing website from the legitimate ones. They have extracted the features automatically without any human intervention by using their own software tool. Based on the selected features they have proposed a set of rules which are used to distinguish a phishing website from the legitimate ones. The dataset used by them collected from PhishTank [23]. From that, the authors have considered 17 features by calculating the frequency of each features. Finally, they have identified "Request URL", "HTTPS and SSL" are more significant and "Disabling Right Click" followed by "URL having @ symbol" are low significant features. They are not reason out why they are only considered 17 features and how they identified those features, among the other features available in the dataset. They are not addressed the group of features which are influenced in phishing attack

In all previous research [17 14 13 24 11] work has been focused on improving the phishing attack classification rate, identify the best classifier to identify phishing attack, address the different types of phishing detection approaches and feature selection of the phishing URL to reduce the dimension of the phishing dataset. But they failed to identify strategies or techniques, cyber-criminals used to mimic URLs to manipulate humans. Features identified in previous research work not only focused on URLs, but also domain-related (i.e. registration, indexing) and content-related (occurrences of other URLs though). Therefore, previous work failed to address how URL-related, domain-related and content-related features are combined to manipulate humans or identify features that are more successful in launching a phishing attack. In most of these previous research work just provided the classification rates as an outcome, but they failed to mention how these classification rates are supported in developing anti-phishing educational interventions. There has been a lack of research reported in the past understanding phishers' strategies of mimicking URLs to leverage phishing attacks through human manipulation, thus using a large dataset (i.e. analysing using ML techniques). Therefore, this research focuses on investigating the feature selection of phishing URLs (Uniform Resource

Locators), aiming to explore the strategies employed by phishers to mimic URLs that can obviously trick people into clicking links or download a malicious IT application (e.g. ransomware attack) that can disturb the normal behaviour of a computer system.

# 3 METHODOLOGY

This research focuses on investigating strategies or techniques that are used by phishers to mimic URLs to leverage phishing attacks through manipulating humans. Figure 1 illustrates how we identify and evaluate the topmost features (techniques to mimic URLs) from the dataset which is downloaded from [1]. This dataset contains 48 features related to the phishing URL and the phishing website. Target class labeled either Legitimate (1) or Phishing (0). Phishing webpage data downloaded from PhishTank[2], OpenPhish[3] and Legitimate ones are downloaded from Alexa[4] and Common Crawl[5]

According to Figure 1, methodology of this research work has two phases. Phase 1 identifies the topmost features that are employed by cyber-criminals to mimic URLs. To identify the topmost features we are assessing their frequency measure. The frequency measure explains how a specific feature influences the target class (Phishing or Legitimate). Higher frequency score feature considered as the most influenced feature in mimic URLs for phishing attacks. To measure the frequency score within the URL feature and target class, we employed Information Gain (IG) and Chi-Squared feature selection methods[26].

The Algorithm 1 describes how to perform Information Gain (IG) and Chi-Squared on the phishing URL features. To calculates the IG for each URL feature, first it calculates the entropy of the phishing URLs' data.

4 '

Entropy [27] is used to measure the impurity, disorder or uncertainty in a bunch of data. As stated in the equation 1 P(x) is simply the frequentest probability of class x. In this experiment, the class x can either be a phishing(1) or legitimate(0). Then the value obtain from the equation 1 represents the frequentest probability of class phishing and legitimate. This entropy value is considered as a parent entropy (Entropy(parent)) during the calculation of Information Gain (IG) on phishing URL features.

$$\text{Entropy} = - \sum P(x) \times \log_2 P(x) \tag{1}$$

We then calculate the IG for each phishing URL feature using the entropy score. According to the equation 2, the entropy value of each phishing URL feature (i.e: Entropy(children))is subtracted from the parent entropy (Entropy(parent)). The resultant value is an IG of the specific phishing URL's feature. The obtained score describes how a specific phishing URL feature influences the identification of the target class (phishing, legitimate). The most influential techniques of manipulating phishing URL are identified based on the IG score features

$$\text{Information gain} = \text{Entropy(parent)} - \text{Weighted auerage} \times \text{Entropy(children)} \tag{2}$$

$$\text{Weighted auerage} = \text{no. of examples in the child node} / \text{total number of example in parent node} \tag{3}$$

Then we perform the calculation of Chi-Squared score for each phishing URL feature based on the equation 4. In the Chi-Squared [28] approach, we compare each phishing URL feature with the target class(phishing). If they are independent, then observed count of the class(phishing) is close to the expected count of the class(phishing).Thus we get a smaller $x^2$ value. On the other hand, if the phishing URL feature is dependent with the target class(phishing), we will get a higher $x^2$ value. We have used this technique to identify the dependency between each URL feature and target class(phishing). Table 1 shows the $x^2$ values for top 20 features. The features present in the Table 1 highly depend on the target class(phishing). Thus, they are selected as an important technique of mimicking URLs for phishing attacks

$$x^2 = \frac{1}{d} \sum_{k=1}^{n} \frac{(O_k - E_k)^2}{E_k} \tag{4}$$

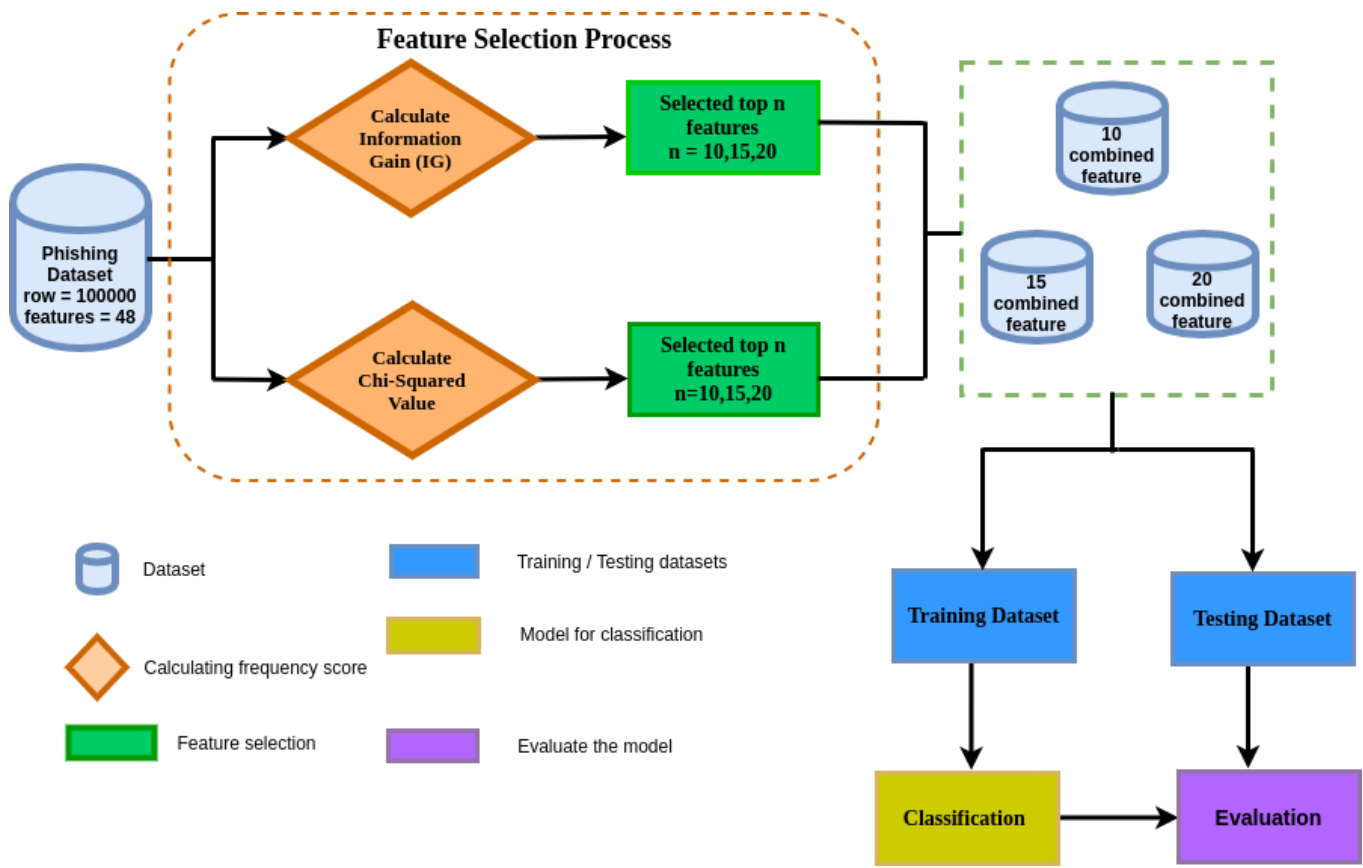

**FIGURE 1** Methodology for Identifying Phishing URL

$O_k$ = number of observations in class k

$E_k$ = number of expected observations in class k

Once we have calculated the IG and Chi-Squared score for each URL feature, we are sorted those scores in descending order. After that we create three combined datasets by combining top n = 10, 15 and 20 features obtained from IG and Chi-Squared. Then we have used these combined datasets for the classification of phishing URLs.

---

**Algorithm 1** Select top n features

---

1: **procedure** SELECTTOPNFEATURES(listoffeatures, n)                          ◁ Step1: select top n features using IG
2:                                                                               ◁ n: 10,15 and 20
3:     igFeatures ► calculateIg(listoffeatures)
4:     sortedIgFeatures ► sort(listoffeatures)
5:     selectedTopnIg › selectfeatures(sortedIGFeatures)          ◁ Step2: select top n features using Chi-Squared
6:
7:     chiFeatures ► calculateChiSquared(listoffeatures)
8:     sortedChiFeatures ► sort(listoffeatures)
9:     selectedTopnChi ► selectfeatures(sortedChiFeatures) ◁ Step3: Create combined dataset by using step1 and step2
10:
11:    combinednfeatures ► combine(selectedTopnIg, selectedTopnChi)
12:    **return** combinednfeatures
13: **end procedure**

---

Al-diabat, Mofleh [13] also use IG and Symmetrical Uncertainty (SU) for feature selection. Authors use the common features obtained from both the methods for classification. Abdelhamid, Neda and Ayesh, Aladdin and Thabtah, Fadi [17] proposed research work, that they use Chi-Squared to identify the dependency among the features and to group those dependent features for classification. However, when perform grouping among the features, we may miss some important influencial features. Therefore, for this research, we have combined the features that are obtained from IG and Chi-Squared. Our intention was to identify the top features(i.e. URL mimicking techniques) that are used by phishers to mimic URLs to leverage phishing attacks through manipulating humans.

In phase 2, we perform a classification to evaluate three combined datasets. In this regard, three combined datasets are guiding one by one into three well known classifiers such as Naive-Bayes, LinearSVC (Support Vector Classifier) and K Nearest Neighbors (KNN). Once the classifiers are trained, they are evaluated against the testing data. The Figure 2 illustrates a comparison based on the performance between the original dataset (for 48 features) and the combined datasets (for 10, 15 and 20 features). The accuracy of the classification using combined datasets is comparatively higher than the original dataset. Based on the accuracy result presented in Table 2 we can conclude that the identified features available in the combined datasets are the most influential techniques employed by cybercriminals' to mimic URLs. In this research, we only focuses on 10, 15 and 20 features as the level of accuracy performed high (almost 100%) for classifying phishing.

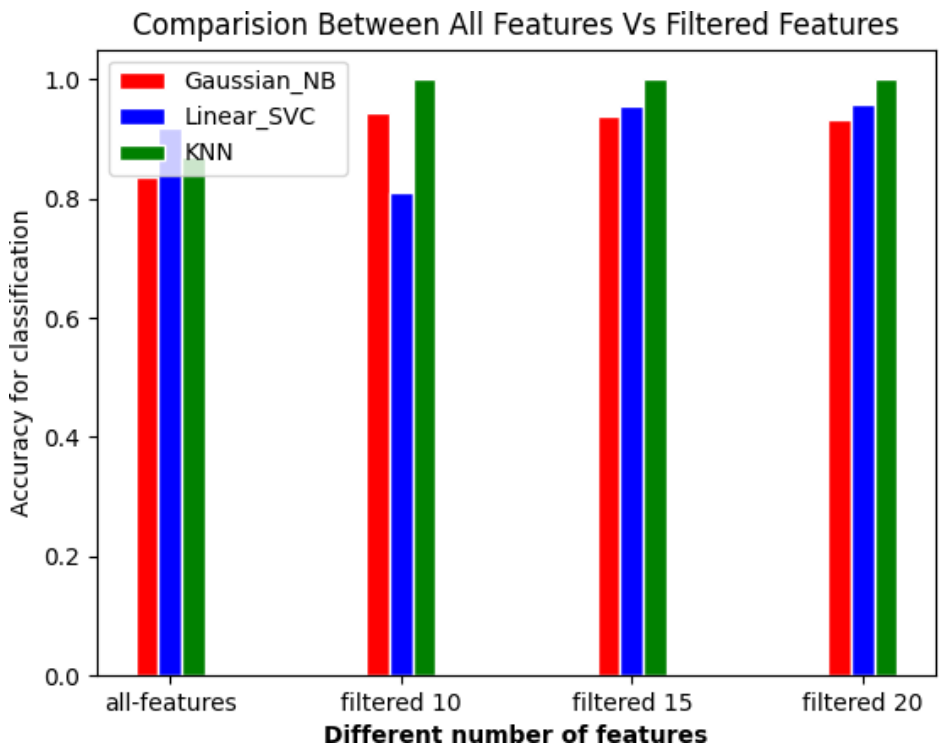

**FIGURE 2** Classification rate for different number of features

## 4 . RESULT

### 4.1 . Dataset

The dataset of phishing websites is downloaded from the [21] site. A total of 48 features is extracted from 5000 phishing webpages and another 5000 legitimate webpages, downloaded from January to May 2015 and from May to June 2017. Target class labeled either Legitimate (1) or Phishing (0). Phishing webpage data downloaded from PhishTank [23], OpenPhish [29] and Legitimate ones are downloaded from Alexa [30] and Common Crawl [31].

### 4.2 . Experimental Setup

Phishing dataset consists of 10000 records. We have split our dataset (70%) for training and (30%) for testing. Then we did the feature selection process by using IG and Chi-Squared as stated in the above sections 3.2 and 3.3. Thereafter, we have performed classification for best 20, 15 and 10 features. Table 2 provides a detailed accuracy rate for different number of features. For top 10 features K-Nearest-Neighbour(KNN) classifier performed well than other two classifiers. When we increase the number of features as 15 and 20 LinearSVC and KNN, are providing best accuracy rates.

### 4.3 . Summary of the Result

In the past literature proposed by Al-diabat, Mofleh [13] they found that SSL-Final-state and URL-of-Anchor are the two top-most phishing website related features. In the Mohammad, Rami M and Thabtah, Fadi and McCluskey, Lee [14] research work they have identified "Request URL", "HTTPS and SSL" are more significant and features in phishing website. Both of them used UCI Phishing dataset [22] for their experiment. This dataset is pretty much old and most of the features relate to phishing websites. But in our experiment we have used new phishing dataset [21] since the data was collected between 2015 to 2017 and most of the features are related to the phishing URL. Among the 48 features, our results revealed that PctNullSelfRedirectHyperlinks, FrequentDomainNameMismatch, SubmitInfoToEmail, PctExtResourceUrls, InsecureForms, ExtMetaScriptLinkRT, PctExtNullSelfRedirectHyperlinksRT, NumDash, IframeOrFrame, NumSensitiveWords, PctExtHyperlinks, NumNumericChars and NumDots are to be the topmost URL mimic techniques. Among these Null Self Redirect Hyperlinks in URL, Number of Dashes in URL, Submit Information to Mail, Insecure Forms, IframeOrFrame and URL Attached with the Number of Sensitive Words are the most recent URL mimicking techniques. These techniques are employed by phishers to mimic URLs to leverage their attacks through manipulating humans. These techniques well influenced in the modern phishing attack because of the growth of heterogeneity of modern devices, the number of users using social media Apps and insecure online shopping or banking applications.

## 5 . DISCUSSION

In this section we are going to discuss how the identifies feature are employed by phishers to mimic URLs to leverage their attacks.

**Null Self Redirect Hyperlinks in URL:** The user hits a link (anchor tag) on a web page, and it opens in a new browser tab. In this stage, chances are high that a hacker might have taken a control over the user's original tab web page. For example attached **target="_blank"** inside the anchor tag like as follows

```
<a href="//fossbytes.com" target="_blank">Fossbytes</a>.
```

If user click on the link it will redirect to the new tab and pointing to another web page. That page is actually a malicious page and it has full control over the previous page's document. The attacker designed that page to look like the original page. Then asking user's login credentials or credit card details. But user likely wouldn't notice this because the redirect happened in the background as shown below in Figure **??**.

```
                    ],
                    "application_context" => [
                        "cancel_url" =>  route('paypal',$payment->id),
                        "return_url" =>  route('paypal',$payment->id)
                    ]
                ];
        $paypal_order = $paypal->execute($order);
        $redirect = collect($paypal_order->result->links)->first(function($link){
            return $link->rel == 'approve';
        });
        $payment->update([
            'payment_id'=> $paypal_order->result->id,
            'redirect'=>$redirect->href??null
        ]);
        // problem is herer
        $tx = $paypal->execute(new OrdersGetRequest($paypal_order->result->id));
```

**FIGURE 3** Null Self Redirect Hyperlinks in URL

**Domain Name Mismatch:** Domain names are hijacked with the intent to steal customers' data and taken out the competitor's website. For example, a domain name info.brienposy.com would be a child domain of brienposy.com, because it appears at the end of the full domain name. But the domain name, brienposy.com.malicious.com not originated from brienposy.com since the parent domain name brienposy.com is on the left side of the full domain name. This URL mimics by the attacker to fool the user that it's from the brienposy.com. This kind of trick used by phishers as a means of trying to convince victims the message or email came from a well-known company. Figure **??** illustrates how the phishers mimic the URL for the website myetherwallet.com as myetherwallel.com

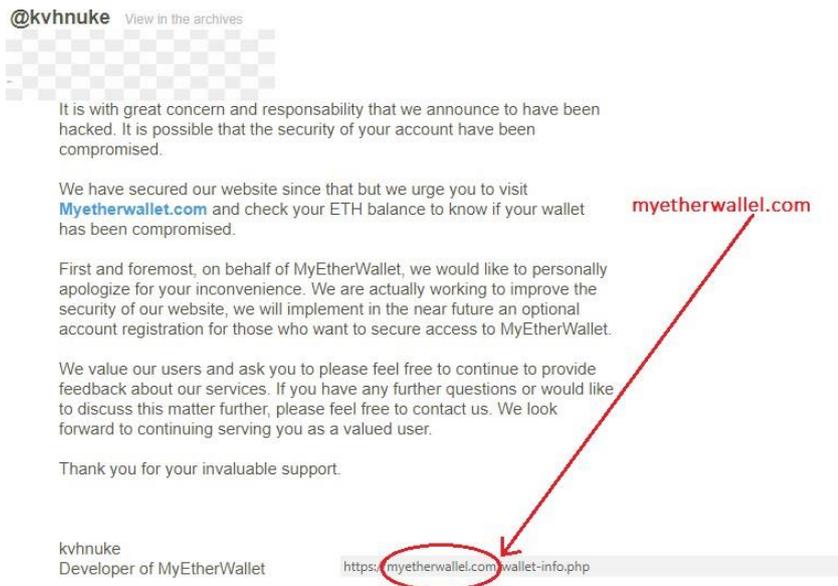

**FIGURE 4** Domain Name Mismatch

**Number of Dashes in URL:** Phishers imitate legitimate domain names by inserting dashes into the URL. An unsuspecting user believes it is legitimate domain name. For example, the phishing domain name http://www.pay-pal.com is imitating PayPal domain name, http://www.paypal.com. However, the use of dashes in a domain name is rarely seen on a legitimate website. This technique can easily deceive users who do not understand the syntax of the URL and cannot tell the different domain name. Figure **??** shows an example of the URL with dashes

**FIGURE 5** Number of Dashes in URL

**Submit Information to Mail:** Phishers create an email which claims user is enticed www.google.com/ConfirmAccount to confirm an ownership of their account. When the user clicks on the URL, a malicious script executes in the background to hijack the user's account information. Then the attacker, monitor the page, hijacks the original password to gain access to secured data of the user's account. For example, Figure 6 shows an email which is mimicked as receive from an www.google.com. It asked their customer to verify the account ownership by clicking on a link (i.e. phishing link) of the confirmation email. For example, www.google.com/ConfirmAccount?Email=hijacker@gmail.com is a kind of phishing link, sending the user's account information to an attacker's email (hijacker@gmail.com). This is another important strategy in email phishing that cyber-criminals used to manipulate humans to disclose their information.

**Insecure Forms:** The webpage contains a form which does not use https, rather it uses http which is insecure. Hackers copied the layout of the login page of the well-known companies like Microsoft, Apple, etc. Then asks users to verify their account or re-set password for security purpose. When the user gives personal information it will send to the hacker's own server or their database instead of Microsoft, Apple, Facebook, etc application's original server. Figure **??** is an example for the fraudulent Microsoft account web page it asks the user to verify their account details, but the URL of the webpage is not related to the Microsoft.

**Number of Dots in URL:** Phishing URL usually has many dots to make users believe that they are genuine page. Chiew, Kang Leng and Choo, Jeffrey Soon-Fatt and Sze, San Nah and Yong, Kelvin SC[32] stated legitimate website has at most five dots in the URL domain while most phishing websites have five or more dots in the URL domain. Phishers use such tricks to obfuscate internet users from perceiving the actual phishing URL. For example the following URL has six dots in the domain name www.network.solutions.com.012892378267.239827432.mobi/login,secure. When an attacker manipulates this kind of URL, usually they ask the user to perform some action like a login. But the user might not notice the number of dots in the domain name. When user click on the link to perform some action from that moment user's account became under the control of the attacker. It is a well-known phishing strategy used by attackers in targeting the domain name based phishing attack.

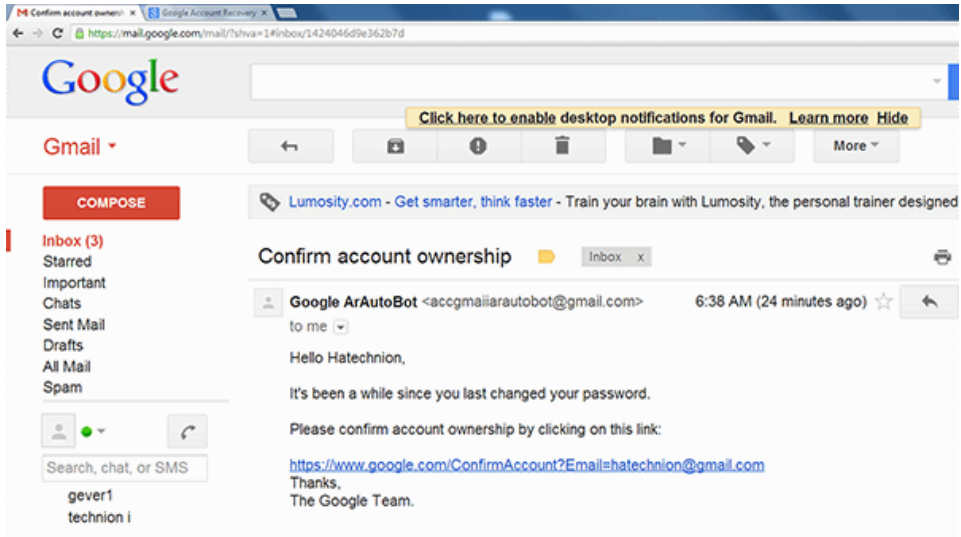

**FIGURE 6** Submit Information to Mail

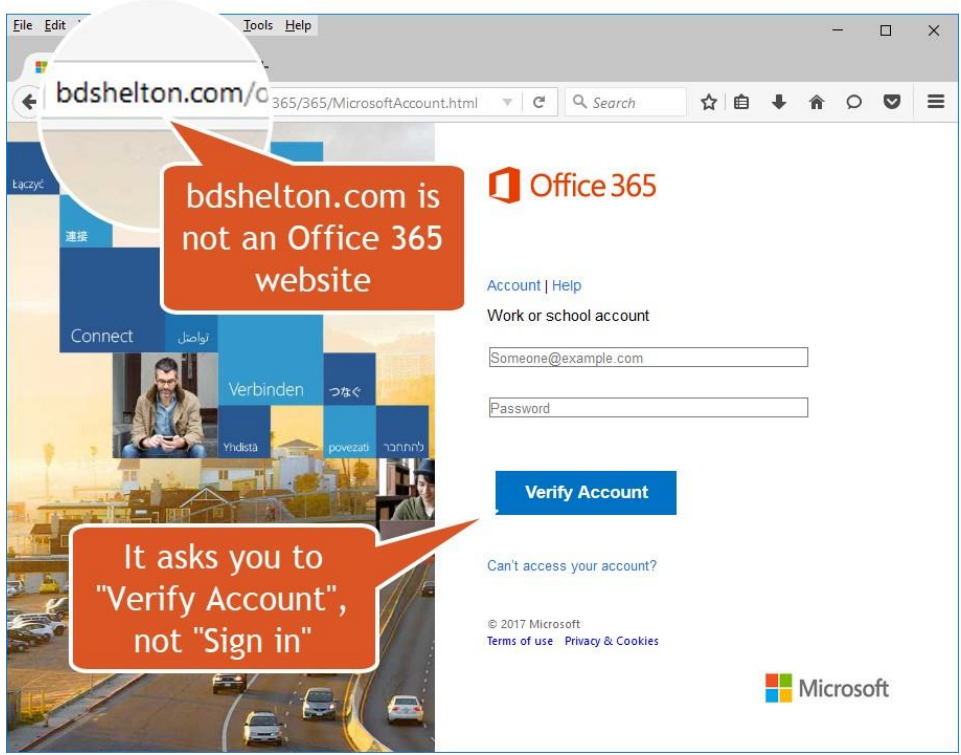

**FIGURE 7** Insecure Forms

**URL Attached with the Number of Sensitive Words:** Phishing URLs attached within the sensitive words to pretend to be legitimate websites. For example, an attacker sends a phishing URL with the sensitive words like secure, account, update, login, sign-in, banking confirm and verify can force users to click on the URL and submit forms with their private information. Figure 8 shows an example for sensitive words in mimicking a URL. An attacker mimics Bank of America security email and create an urgency by using the word "We detect unusual activity in your debit card". It's another kind of strategy based on the sentiment analysis of the user.

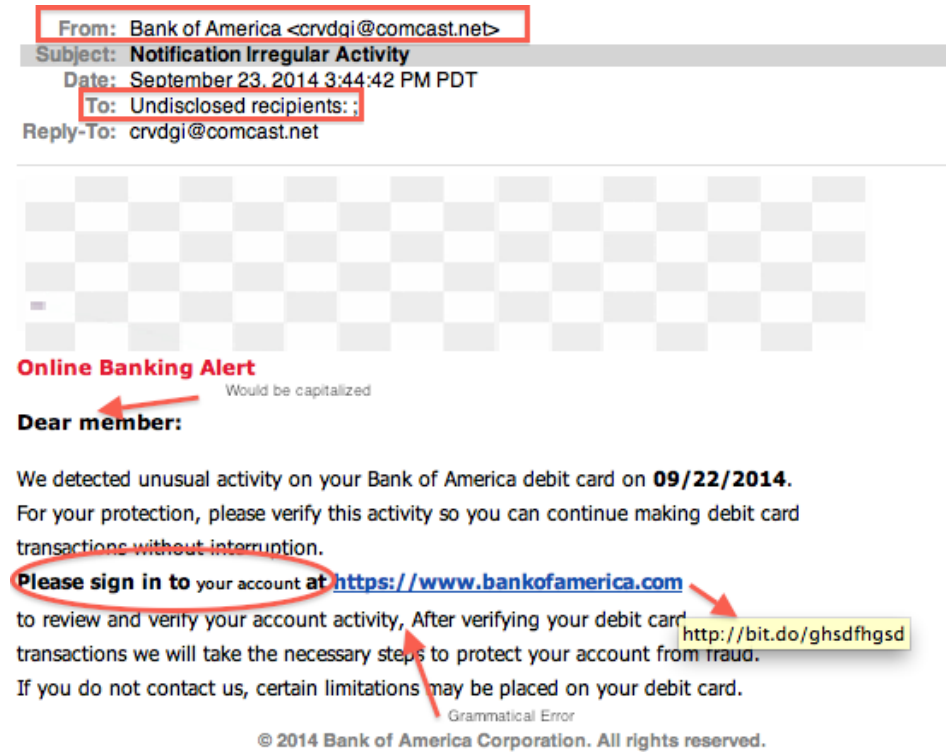

**FIGURE 8** Number of sensitive words

**IFrame or Frame:** The IFrame is an HTML document embedded inside another HTML document. IFrame injection is known as cross-site scripting attack [33]. It consists of one or more IFrame tags that have been inserted into a page or post's content. Which is typically downloaded an executable program or conducts other actions that compromise the victim's computer. For example, if the URL of the webpage like as follows http://www.google.com/index.php?ParamUrl=robots.txtParamWidth=250ParamHeight=250 then this a sign of IFrame injection. Since an attacker can run a malicious script by closing the ifram tag as shown below http://www.google.com/index.php?ParamUrl=robots.txtParamWidth=250ParamHeight=250 "></iframe><a href="javascript:void(document.cookie="authorization=true")"></a>. This will result to change the current cookie parameter authorization=false to authorization=true, then a malicious user will be able to gain access to the sensitive information of the user.

In addition to that, we also identified the co-relations among these techniques. The co-relation matrix in Figure 9 illustrates that when an attacker sends an email with a mismatch domain named URL, it redirects into the attacker's website through the embedded hyperlink. Since the techniques PctExtNullSelfRedirectHyperlinksRT and FrequentDomainNameMismatch show at 0.6 (60%)co-relation. It also showed 30% co-relation between NumDots and NumSensitiveWords which means when an attacker manipulate the URLs with unwanted dots in the domain name mostly attached that URLs with the sensitive words/text. This is another strategy employed by the attackers to steal user's information.

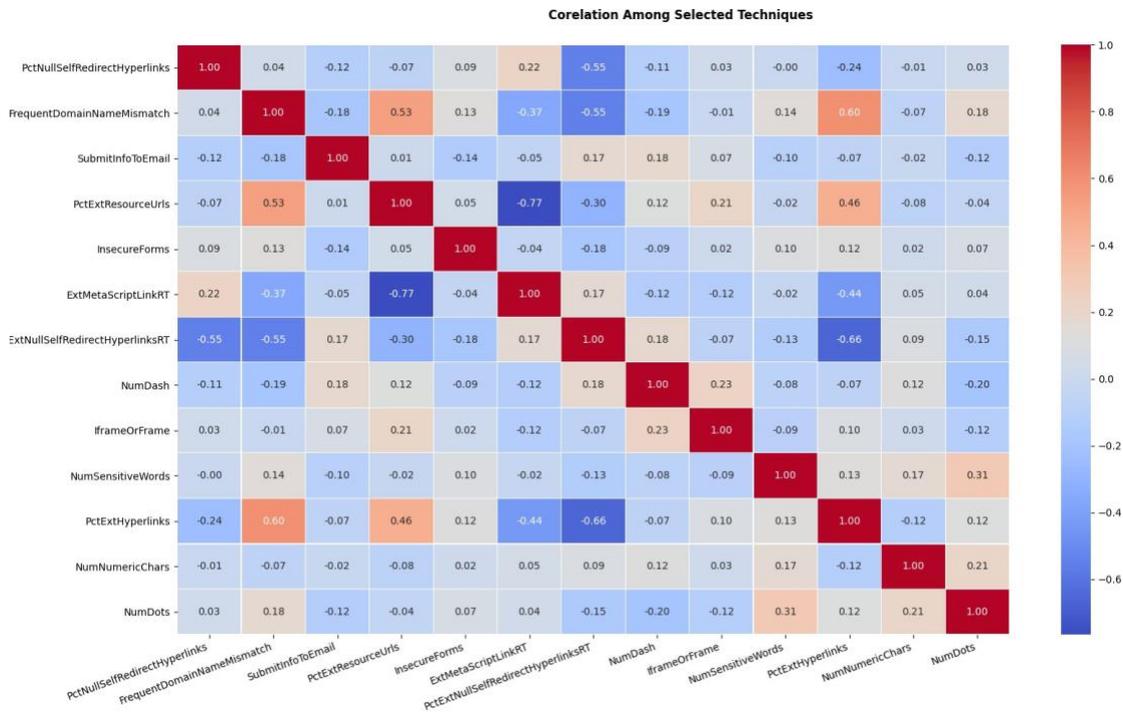

**FIGURE 9** Co-relation among selected techniques

# 6 . CONCLUSIONS

Investigating techniques used by phishers to trick people into disclosing their credentials are vital. Therefore, this research investigates the feature selection of phishing URLs (Uniform Resource Locators), aiming to explore the techniques employed by phishers to mimic phishing URLs to lure people to do malicious tasks such as clicking on fake links. We employed the feature selection method, namely Information Gain (IG) and Chi-Squared, in ML, through a phishing dataset [21]. This dataset contains 48 features extracted from 5000 phishing URLs (PhishTank, OpenPhish) and 5000 legitimate URLs (Alexa, Common Crawl), downloaded from January to May 2015 and from May to June 2017. Our results revealed that the top 10 techniques phishers employed to mimic URL to leverage their attacks through manipulating humans. Among them, Null Self Redirect Hyperlinks in URL and Domain Name Mismatch are the most often used techniques to mimic URLs that can trick people to perform various malicious functions. Furthermore, identify phishing techniques can improve the phishing detection solutions. Improving phishing detection accuracy when a long-term benign domain decides to begin carrying out malicious phishing activity; and reducing the number of false positives for benign domains that are running for a very short period of time. We believe this research will improve the anti-phishing training in terms of gamification to train user against phishing attacks. For example, cyber security educational interventions can be designed and developed understanding the strategies or techniques employed by cyber-criminals.

## References

1. Web Services. https://en.wikipedia.org/wiki/Social_networking_service; 2019.

2. Abdulhameed M, Arachchilage NAG. A Model for the Adoption Process of Information System Security Innovations in Organisations: A Theoretical Perspective. *27th Australasian Conference on Information Systems (ACIS), University of Wollongong, Australia* 2016: 12.

3. Arachchilage NAG, Love S. Security awareness of computer users: A phishing threat avoidance perspective. *Computers in Human Behavior* 2014; 38: 304–312.

4. Benavides E, Fuertes W, Sanchez S, Sanchez M. Classification of Phishing Attack Solutions by Employing Deep Learning Techniques: A Systematic Literature Review. In: Springer. 2020 (pp. 51–64).

5. Phishing. https://www.imperva.com/learn/application-security/phishing-attack-scam//; 2018.

6. Arachchilage NAG, Love S, Beznosov K. Phishing threat avoidance behaviour: An empirical investigation. *Computers in Human Behavior* 2016; 60: 185–197.

7. Gupta B, Arachchilage NA, Psannis KE. Defending against phishing attacks: taxonomy of methods, current issues and future directions. *Telecommunication Systems* 2018; 67(2): 247–267.

8. Dixon M, Gamagedara Arachchilage NA, Nicholson J. Engaging Users with Educational Games: The Case of Phishing. In: ACM. ; 2019: LBW0265.

9. PATYAL M, Sampalli S, Ye Q, RAHMAN M. Multi-layered defense architecture against ransomware. *International Journal of Business and Cyber Security* 2017; 1(2).

10. Arachchilage G, Asanka N. *Security awareness of computer users: A game based learning approach*. PhD thesis. Brunel University, School of Information Systems, Computing and Mathematics, 2012.

11. Sahingoz OK, Buber E, Demir O, Diri B. Machine learning based phishing detection from URLs. *Expert Systems with Applications* 2019; 117: 345–357.

12. OpenUrl. https://blog.qualys.com/securitylabs/2016/01/07/open-redirection-a-simple-vulnerability-threatens-your-web-applications; 2019.

13. Al-diabat M. Detection and Prediction of Phishing Websites using Classification Mining Techniques. *International Journal of Computer Applications* 2016; 147(5).

14. Mohammad RM, Thabtah F, McCluskey L. An assessment of features related to phishing websites using an automated technique. In: IEEE. ; 2012: 492–497.

15. Sharifi M, Siadati SH. A phishing sites blacklist generator. In: IEEE. ; 2008: 840–843.

16. Ali W. Phishing Website Detection based on Supervised Machine Learning with Wrapper Features Selection. *International Journal of Advanced Computer Science and Applications* 2017; 8(9): 72–78.

17. Abdelhamid N, Ayesh A, Thabtah F. Phishing detection based associative classification data mining. *Expert Systems with Applications* 2014; 41(13): 5948–5959.

18. Mohammad RM, Thabtah F, McCluskey L. Predicting phishing websites based on self-structuring neural network. *Neural Computing and Applications* 2014; 25(2): 443–458.

19. Fernando M, Arachchilage NAG. Why Johnny can't rely on anti-phishing educational interventions to protect himself against contemporary phishing attacks?. *30th Australasian Conference on Information Systems (ACIS), Perth Western Australia* 2019: 12.

20. James J, Sandhya L, Thomas C. Detection of phishing URLs using machine learning techniques. In: IEEE. ; 2013: 304–309.

21. Phishing Dataset for Machine Learning. https://data.mendeley.com/datasets/h3cgnj8hft/1; January to May 2015 and from May to June 2017.

22. UCI Dataset. https://archive.ics.uci.edu/ml/datasets/phishing+websites; 2015 March 26.

23. Phishtank. https://phishtank.com/; 2006.

24. Jain AK, Gupta BB. Phishing detection: analysis of visual similarity based approaches. *Security and Communication Networks* 2017; 2017.

25. Yahoo. https://en.wikipedia.org/wiki/Yahoo!_Directory/; 2018.

26. Al-Harbi O. A Comparative Study of Feature Selection Methods for Dialectal Arabic Sentiment Classification Using Support Vector Machine. *arXiv preprint arXiv:1902.06242* 2019.

27. Entropy. https://en.wikipedia.org/wiki/Entropy; 2020.

28. Chi-Squared. https://towardsdatascience.com/chi-square-test-for-feature-selection-in-machine-learning-206b1f0b8223; 2019.

29. Openphish. https://openphish.com/; 2014.

30. Alexa. https://www.alexa.com/topsites/category/Computers/Internet/On_the_Web/Web_Applications/Databases; 1996-2019.

31. CommonCrawl. https://commoncrawl.org/; 2010.

32. Chiew KL, Choo JSF, Sze SN, Yong KS. Leverage website favicon to detect phishing websites. *Security and Communication Networks* 2018; 2018.

33. Cross Site Scripting Attack. https://www.acunetix.com/websitesecurity/cross-site-scripting/; 2020.

| URL Feature | $x^2$ ±0.01 |
|---|---|
| PctExtNullSelfRedirectHyperlinksRT | 3028.93 |
| FrequentDomainNameMismatch | 1971.20 |
| NumDash | 1138.26 |
| SubmitInfoToEmail | 1027.15 |
| PctNullSelfRedirectHyperlinks | 944.58 |
| InsecureForms | 745.96 |
| NumDots | 682.73 |
| PctExtHyperlinks | 550.03 |
| NumSensitiveWords | 505.98 |
| IframeOrFrame | 399.65 |
| PathLevel | 363.94 |
| AbnormalExtFormActionR | 221.52 |
| UrlLengthRT | 199.51 |
| HostnameLength | 194.87 |
| NumDashInHostname | 168.57 |
| NumQueryComponents | 154.48 |
| EmbeddedBrandName | 152.04 |
| AbnormalFormAction | 135.22 |
| IpAddress | 120.64 |
| DomainInPaths | 106.73 |

**TABLE 1** Chi-Squared value for URL features.

| Classifiers | Number of features | | | |
|---|---|---|---|---|
| | 48 | 10 | 15 | 20 |
| Naive-Bayes | 0.837 | 0.945 | 0.938 | 0.931 |
| LinearSVC | 0.918 | 0.809 | 0.954 | 0.957 |
| K-Nearest-Neighbours | 0.871 | 1.0 | 1.0 | 1.0 |

**TABLE 2** .Classification rate for different number of features